\documentclass{aastex701}

\usepackage{graphicx}
\usepackage{tabularx}
\usepackage{array}
\usepackage{multirow}
\usepackage{chemformula}
\usepackage{amsmath}
\usepackage{appendix}
\usepackage{natbib}
\usepackage{CJK}
\usepackage{fontspec}
\usepackage{float}
\bibpunct{(}{)}{;}{a}{}{,}

\usepackage{hyperref}
\hypersetup{colorlinks=true, linkcolor=red, citecolor=blue}

\graphicspath{{figures/}}

\begin{document}

\title{Diversity in planetary architectures from pebble accretion: Water delivery to the habitable zone with pebble snow} 

\correspondingauthor{Sean McCloat}
\email{spmccloat@gmail.com}

\author[0000-0002-3654-9818, gname=Sean, sname=McCloat]{Sean McCloat}
\affiliation{Department of Space Studies, School of Aerospace Science, 4149 University Avenue, Stop 9008, Grand Forks, ND 58202, USA}
\email{spmccloat@gmail.com}

\author[0000-0002-1078-9493]{Gijs D. Mulders}
\affil{Instituto de Astrof\'isica, Pontificia Universidad Cat\'olica de Chile, Av. Vicu\~na Mackenna 4860, 7820436 Macul, Santiago, Chile}
\email{gdmulders@gmail.com}

\author[0000-0003-1982-0474, gname=Sherry, sname=Fieber-Beyer]{Sherry Fieber-Beyer}
\affiliation{Department of Space Studies, School of Aerospace Science, 4149 University Avenue, Stop 9008, Grand Forks, ND 58202, USA}
\email{sherry.fieber.beyer@und.edu}

\begin{abstract}
"Pebble snow" describes a planet formation mechanism where icy pebbles in the outer disk reach inner planet embryos as the water ice line evolves inward. We model the effects pebble snow has on sculpting planetary system architectures by developing "The PPOLs Model"\footnote{The code, docs, and tutorial are publicly available at https://doi.org/10.5281/zenodo.17137567}. The model is capable of growing any number of protoplanet seed masses by pebble accretion simultaneously and accounts for differences in rocky and icy pebble composition, the filtering of pebbles by other protoplanets, the pebble isolation mass, and a self-consistently evolving snow line. The growth and bulk composition are recorded across a grid of protoplanetary disks with stellar masses ranging from 0.125 - 2.0${M_{\odot}}$ (M to A stars) and disk masses ranging from 1 - 40\% of the stellar mass. Three system architectures emerge following a low-, mid-, and high-disk mass fraction that remains consistent across stellar mass. The low-mass architecture is the only one to yield short period Mars-Earth mass cores with bulk water content spanning orders of magnitude and may be prelude to observed "peas in a pod" systems. The high-mass architecture produces proto-gas giant cores in the outer disk. The middle-mass architecture produces a bimodal peak in mass within a system, with the outer protoplanet mass at the snow line growing to an order of magnitude larger, resembling the Solar System. Solar system-like architectures appear for a small range of initial disk masses around F and G stars, but are not a common feature around K and M stars.
\end{abstract}

\section{Introduction}

Confirmed exoplanets now number in the thousands and although the parameter space of true solar system analogs remains on the edge of observational capability, efforts to distinguish patterns in planetary system architecture are underway. One approach is to explore outcomes of planet formation mechanisms that lead to the planetary systems observed today, and determine how modes of formation may preferentially form Earth-like planets over gas giant planets like Jupiter, water-rich planets over dry planets, or combinations in between. 

Pebble accretion \citep{ormel_effect_2010, lambrechts_rapid_2012} is a mode of planet formation developed to explain how protoplanetary cores in the outer (proto)solar system could grow into gas and ice giants within the lifetime of the protoplanetary disk \citep{pollack_formation_1996, ida_toward_2004}. The theory is supported by observations of protoplanetary disks revealing ubiquitous (drift-dominated) dust/cm solids \citep{andrews_mass_2013, perez_grain_2015, ormel_emerging_2017, banzatti_hints_2020}. In particular, the DSHARP survey \citep{andrews_disk_2018} has revealed distinct rings of concentrated dust are a common feature in disks, and aligns well with theoretical expectations that pebbles would collect in such substructures \citep{dullemond_disk_2018, birnstiel_disk_2018}.

Pebble accretion begins when protoplanetary dust coagulates into millimeter-centimeter grains that experience aerodynamic drag from the protoplanetary gas and thereby drift inward toward the central star. Pebbles drifting past a planetesimal are subject to both gravitational focusing and aerodynamic drag that increases accretion efficiency of the seed mass beyond gravitational focusing alone. The sustained inward flux of pebbles may explain the mass source for compact super-Earth systems like the TRAPPIST-1 planets \citep{schoonenberg_pebble-driven_2019, liu_super-earth_2019}.

However, the pebble flux reaching inner protoplanets is reduced as drifting pebbles accrete onto outer protoplanets; the flux can be quenched entirely if an outer seed mass reaches the pebble isolation mass \citep{bitsch_pebble-isolation_2018}. Here, we distinguish between filtering and the pebble isolation mass: we define the former as the cumulative reduction in pebble flux as pebbles are accreted, while the latter as a threshold effect in which the protoplanet mass is sufficient to create a pressure bump in the disk gas and the pebble flux is halted. Observed substructures among protoplaneatry disks are suspected to correspond to these or similar planet formation mechanisms \citep{stammler_dsharp_2019, jiang__efficient_2023}.

A major divider of disk properties is the water ice line, or snow line, which separates the disk into an inner zone where water ice sublimates into gas and an outer zone where the water ice can accrete onto growing protoplanetary cores \citep{morbidelli_fossilized_2016, drazkowska_planetesimal_2017}. These conditions also determine the composition of pebbles as they coagulate out of the locally available solid material, and therefore divides regions where rocky or water-rich pebbles exist. Whether or not a planet can bear significant (e.g. several percent bulk mass) water content despite forming interior to the snow line is a major question in planetary science. The simple observation that Earth is located interior to the snow line poses questions for the source of Earth's water \citep{mottl_water_2007, meech_origin_2020}. The debate on the potential existence of "water worlds" among the abundant compact exoplanet population \citep{luque_density_2022, rogers_conclusive_2023} further highlights the need to explore mechanisms that can deliver and modulate both dry and hydrated mass to the inner disk.

One scenario for modulating the composition of solids of the inner disk invokes inward evolution of the snow line thereby allowing icy pebbles to sweep over otherwise rocky and dry protoplanet masses; we refer to this scenario as "pebble snow".\footnote{This scenario is summarized and referenced by Raymond et al. in a chapter from \citep{meadows2020planetary}} \cite{sato_water_2016} and \cite{ida_water_2019} model the ability for icy pebbles aggregating in the outer disk to accrete onto cores with the current mass and location of Venus, Earth, and Mars, as individual and isolated cores. They find that that the water mass fraction of an embryo is regulated by the icy dust (not pebbles) remaining in the outer disk at the time the snow line sweeps past it – for example, preserving the overwhelmingly dry character of Earth (0.023\%, or at most, 1\%) occurs when the snow line arrives after $\sim$4 Myr for disks 100 au in size, and never for disks 300 au in size. Among other considerations, they note that future work in this direction would benefit from accounting for the fragmentation of icy pebbles, the effect of one or multiple outer giant planet cores, and an evolution of the snow line that is consistent with other factors of disk evolution.\footnote{We use the term "evolution" referring to the snow line and changes in the protoplanetary disk to avoid confusion with "migration" referring to protoplanets, or "drifting" referring to pebbles.} 

\cite{mulders_why_2021} explore the effect of an outer protoplanet growing at 5 au (a Jupiter analog) on an inner protoplanet at 0.3 au (a super-Earth analog) by accounting for both filtering, the pebble isolation mass threshold, and a fixed snow line. They find a difference in outcomes when both filtering and the snow line are considered for different stellar and disk mass combinations: super-Earths are better able to grow around M dwarf stars compared to higher mass stars. There is less dust mass around lower mass stars from which to efficiently grow the outer protoplanet, and therefore the outer protoplanet cannot quench the inner seed growth. This may explain the observed abundance of super-Earths around M dwarf stars.

Here, we expand on this framework by advancing a pebble accretion model that grows protoplanet seed masses during the gas phase of the disk and evolves a snow line self-consistently with changes in the protoplanetary disk. This model grows the protoplanet seed masses simultaneously (instead of one at a time) and accounts for pebble filtering by outer seed masses, pebble isolation mass, and pebble drift and fragmentation. There are several simplifying assumptions made in this model, but the trade-off is the ability to run a grid of models each with 100 (or more) seed masses in less than an hour with a personal computer. By doing so, we explore the system-level impact of "pebble snow" mechanisms on the planetary system architectures up to the point of gas dissipation for a range of stellar mass / disk mass combinations. This paper serves as an introduction of the model and explores how these mechanisms can separate the types of planets and architectures that emerge, with particular focus on the distribution of water content into the inner disk, as well as in the conservative habitable zones of different stars.

\section{Method}
\subsection{Pebble Accretion Model}
In contrast to earlier studies on pebble accretion, we do not develop new expressions for the coagulation, fragmentation, and drift of pebbles. Instead, we use the pebble coagulation/drift and disk model of \textit{pebble-predictor} \citep{drazkowska_how_2021}\footnote{The original code for pebble-predictor is available: https://zenodo.org/badge/latestdoi/300679267} and the pebble accretion efficiency recipes (\textit{epsilon}) from \cite{liu_catching_2018} and \cite{ormel_catching_2018}\footnote{The original "epsilon" code is available on Chris Ormel's website: https://staff.fnwi.uva.nl/c.w.ormel/software.html} to describe pebble accretion and apply this across a range of star/disk combinations. Key changes in the action of pebble accretion that depend on being in the 2D or 3D regime, or the Bondi and Hill encounters, are handled by these prior models. We direct interested readers to those original papers for complete descriptions of the model, but describe essential relevant points here. Below we describe how these two are combined and adapted to explore system architectures. We name the model "The PPOLs Model" after a combination of these two prior models, "PP" for \textit{pebble-predictor} and "OL" from Ormel \& Liu.

 The PPOLs Model builds on this foundation to include:

\begin{enumerate}
    \item disk mass scaled by stellar mass (the disk mass fraction, $f_{\text{disk}}$)
    \item simultaneous growth of seed masses that can vary in location, initial mass, and formation time
    \item pebble filtering and pebble isolation mass
    \item a snow line parameterized by disk conditions that self-consistently evolves with the disk
    \item bifurcated pebble composition between rocky and icy pebbles and tracking of dry or hydrated accreted mass
\end{enumerate}

\textit{Pebble-predictor} takes as input a stellar mass ($M_\text{{star}}$), total solid mass of disk ($M_\text{{solids}}$), dust-to-gas mass fraction ($f_{DG}$), gas and dust surface density profiles($\Sigma_g$, $\Sigma_d$), temperature profile, turbulence described via $\alpha$-viscosity prescription \citep{lynden-bell_evolution_1974}, pebble fragmentation velocity ($v_\text{{frag}}$), and pebble internal density ($\rho_\text{peb}$) for a user-specified disk radius and lifetime. The two outputs are the flux-averaged Stokes number ($St$) and pebble mass flux ($f_\text{{peb}}$) at all locations and times of the user-specified disk. PP produces a single representative Stokes number for the population of pebbles and is presented by \cite{drazkowska_how_2021} as a simplified alternative to other pebble-formation models that calculate a distribution of pebble sizes. The results of their single-size approach are benchmarked to full size distributions and shown in Fig. 7 of that paper.

The solid (dust) disk mass depletes over time as as pebbles coagulate from the dust; further, the inward pebble flux also changes with time and location as pebbles drift or accrete. The decrease of dust mass based on the coagulation and flux of pebbles is tracked for each time step by \textit{pebble-predictor}. However, an important simplification is the gas disk does not evolve and remains constant over the lifetime of the model. The temperature profile is also parameterized as a simple power law instead of accounting for gas accretion or viscous heating. The implications of this are discussed below.

The pebble accretion recipes from Ormel \& Liu (2018)\footnote{the package is named "epsilon", derived from the variable parameter used for pebble accretion efficiency} takes as input a pebble Stokes number and a planetesimal seed mass to yield the the fraction of pebbles that will accrete onto it using analytical fits derived from three-dimensional n-body simulations.  Additional required input includes turbulence (via $\alpha$-viscosity), a gas pressure gradient, and gas disk aspect ratio. The recipes provide flexibility for users to specify different pebble accretion regimes, i.e. from 3D to 2D, or ballistic or settling, or combinations thereof.

The two models are complementary: with one set of disk properties, \textit{pebble-predictor} yields a representative Stokes number and pebble mass flux. The same disk properties, a planetesimal seed mass, and the outputs of \textit{pebble-predictor} are used by "epsilon" to determine the fraction of mass flux that accretes onto the seed mass. The two models allow disk properties to yield consistent pebble properties, available mass, and accretion efficiencies to model the growth history of a planetesimal. The combination is benchmarked in \cite{drazkowska_how_2021} against more complex (and computationally expensive) pebble formation models by growing single-planet systems. The combination is also employed by \cite{mulders_why_2021} to explore the effects of stellar mass on planet formation in two-planet systems.

\subsection{Disk Model}
The PPOLs Model begins with the central stellar mass, which determines the overall disk mass according to: 

    \begin{equation}
    M_\text{{disk}} = \frac{M_\text{{star}}}{M_{\odot}}\:\text{$f_\text{{disk}}$} \,,
   \end{equation}

where $f_\text{{disk}}$ is the disk mass fraction, i.e. the total mass of the disk (gas + solids) as fraction of the stellar mass. This parametrization allows simple and consistent modification of the protoplanetary disk with stellar mass. Alternatively, the disk mass can be explicitly set by the user based on total solid mass in the disk. The mass of the disk is a primary determinant for how much material is available to form planets at all. Observations of protoplanetary disks show scatter with stellar mass \citep{pascucci_steeper_2016, tychoniec_dust_2020}, – further complicated by uncertainty regarding planet formation onset. We explore a wide range of disk mass fractions ($f_\text{{disk}}$ = 1 - 45\%  of stellar mass) to probe limits of the mechanisms at play.

The gas and dust surface density distribution is set by:

   \begin{equation}
        \Sigma_{gas} = 
        \frac{M_\text{{disk}}}{2\pi {r_{c}}^2} \: \bigg(\frac{r}{r_{c}}\bigg)^{-1}  \:e^{-{r}/{r_{c}}} \,,
   \end{equation}

and

   \begin{equation}
        \Sigma_{0,\: \text{solid}} = Z_{0} \: \Sigma_{\text{gas}} \,,
   \end{equation}
   
where $r_{c}$ is the critical radius beyond which the surface density decreases exponentially, representative of flared disks \citep[e.g.][]{kenyon_spectral_1987, smith_discovery_2005, wolff_anatomy_2021}. Although the boundaries of the disk are set to 0.01 - 1000 au, the critical radius describes the effective surface density distribution with rapidly diminishing material for disk radii >200 au. The solid/dust surface density is then characterized by $Z_{0}$, the initial solid-to-gas mass ratio.

The temperature profile is a power law that scales with stellar mass:
\begin{equation}
        T(r) = 280\:  \bigg(\frac{M_{\text{star}}}{M_{\odot}}\bigg)^{1/2} \: \bigg(\frac{r}{\text{au}}\bigg)^{-1/2} \quad\text{K}.
   \end{equation}

This power law profile is observationally supported by analysis of HL Tau \citep{ueda_multi-wavelength_2025}, who find the temperature in the optically thick disk decreases $\propto r^{-0.51}$. However, analytic considerations \citep[e.g.][]{hayashi_structure_1981} suggest this power law is appropriate for optically thin disks. \citet{ida_radial_2016} propose a temperature profile valid in the optically thick regime that accounts for viscous and irradiation heating mechanisms and may be more applicable. We run a series of models using this alternate temperature profile and present those results in Appendix \ref{apx-temp}.

\subsection{Snow line}
\begin{figure}
\centering
\includegraphics[width=0.7 \linewidth]{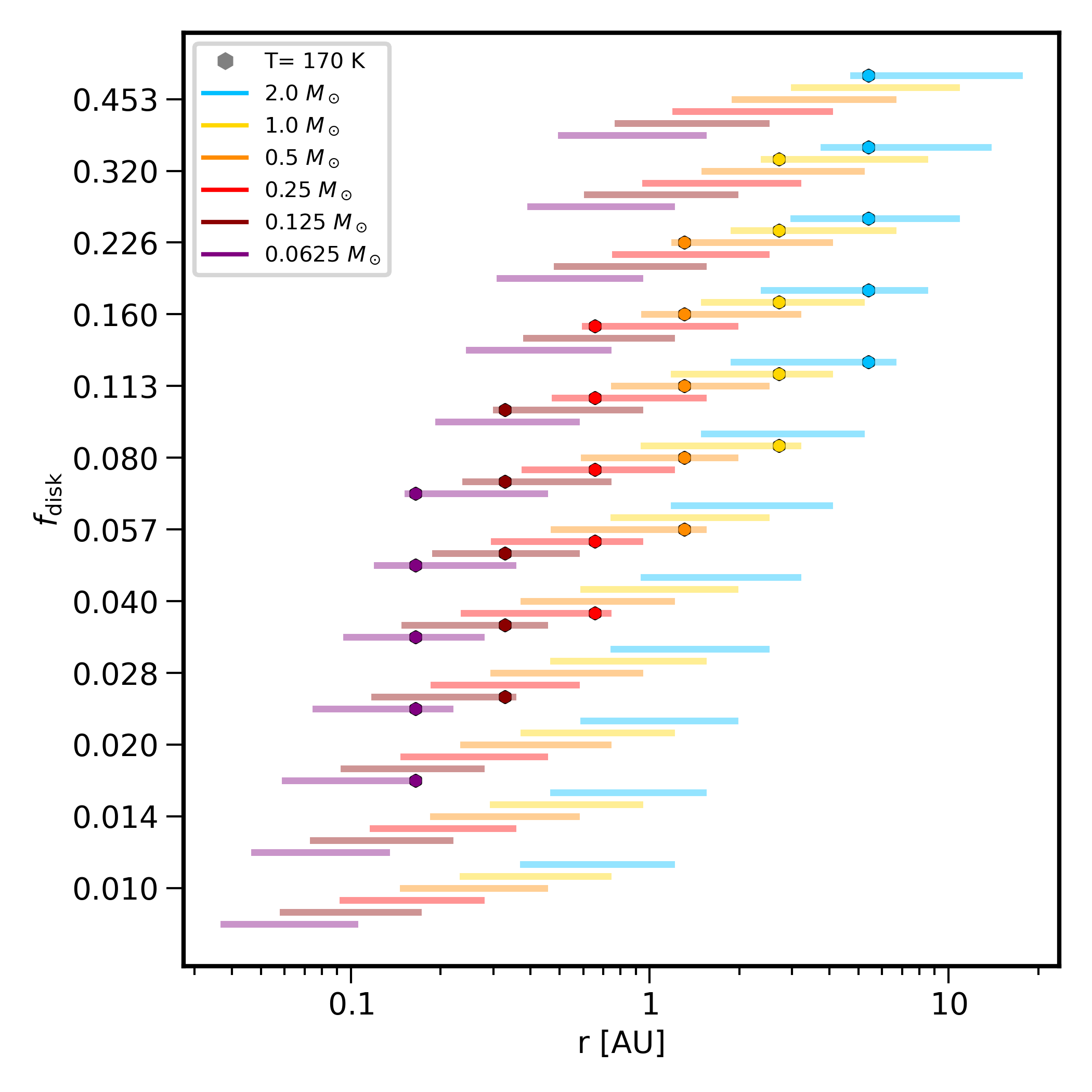}
    \caption{Comparison of the snow line range of this model(bars) compared to disk location where T=170 K (hexagons), for each stellar mass (bar color). All snow lines start at the greater distance and move inward. Markers are placed only where the snow line range overlaps with T = 170 K. The disk temperature profile is a simple function of stellar mass, which causes the temperature marker to appear at the same location per stellar mass. The snow line range changes with stellar mass and disk mass, as parameterized by Eq. \ref{RSL}.}
    \label{FigSLRange}
\end{figure}

A snow line is adapted from Eq. 8 in \cite{savvidou_influence_2020}:

\begin{equation}\label{RSL}
    r_{SL} =
    4.74 \: \bigg(\frac{\alpha}{10^{-2}}\bigg)^{0.61}  
    \bigg(\frac{\Sigma_{0,gas}}{1000\:gcm^-2}\bigg)^{0.704}
    \left(\frac{f_{DG}}{0.01}\right)^{0.37} \quad \text{au} \, ,
\end{equation}

where $\Sigma_{0,gas}$ is the initial gas surface density at 1 au. This parametrization meshes neatly with the disk model already described and allows for self-consistent evolution of the snow line as the disk evolves: \textit{pebble-predictor} tracks the conversion of dust to pebbles which decreases $f_{DG}$. The PPOLs Model tracks this decrease and recalculates $r_{SL}$  at each time step. Figure \ref{FigSLRange} compares the location of $r_{SL}$ in each stellar mass/disk mass combination to the disk location corresponding to T = 170 K. These two are are largely consistent; the disk temperature is a function only of stellar mass, and remains at constant distance for different disk mass fractions. The $r_{SL}$ used in the model is a function of the disk mass (via $\Sigma_{0,\text{gas}}$), and moves inward for lower disk mass fractions.

\cite{savvidou_influence_2020} derive that expression by exploring the feedback between disk temperature and dust grain opacity in equilibrium disk models, and are able to describe the location of the snow line using the $\alpha$-viscosity and initial surface gas density at 1 au. We scale the gas surface density in our expression to account for different parameterizations of the gas surface density between our disk models and yield a snow line consistent with expectations that the snow line in the solar system began at 2 - 3 au \citep{martin_evolution_2012}.

Following \cite{mulders_why_2021}, we assume the snow line defines several changes in the disk, seed masses, and pebble properties. First, we assume the dust surface density is cut in half inside the snow line via sublimation of volatiles. The material composition is bifurcated between rocky material inside to a 50-50 rocky-ice mix outside the snow line. Initial seed mass composition reflects \textit{in situ} formation, such that seed masses inside (outside) the snow line have water mass fractions of 0 (0.5). Similarly, pebbles assume the same bifurcation, but icy pebbles that drift across the snow line lose their volatiles via sublimation and become rocky. Finally, we assume the fragmentation velocity ($v_\text{{frag}}$) of rocky vs icy pebbles changes from 100 cm/s to 1000 cm/s. The fragmentation velocity describes the maximum collisional velocity a pebble can survive before fragmenting \citep{guttler_outcome_2010, zsom_outcome_2010}. A higher $v_\text{{frag}}$ indicates icy pebbles may be stickier or sturdier and coagulate to larger sizes, thus increasing the mass per pebble accreted over their rocky counterparts (but see Sect. \ref{discussion}).

We use a single value for the alpha-viscosity of $10^{-3}$ that is consistent with both prior theoretical considerations as well as observations. \citet{savvidou_influence_2020} use a range of alpha values from $10^{-4}$ - $10^{-2}$. \citet{villenave_turbulence_2025} present uniform analysis for a survey of disk observations and estimate alpha based on dust disk scale heights for 23 of 33 disks in their sample. They find a range of turbulent alpha from $3\times10^{-4}$ - $2\times10^{-2}$, with a median value of $3\times10^{-3}$, as well as a comparable fragmentation (vertical) alpha median value of $1.1\times10^{-3}$. To confine the present analysis, we use a single alpha-viscosity for all models, and we present results for a range of alpha for a 1 $M_{\odot}$ star at a typical low, medium, and high disk mass in Appendix \ref{apx-alpha}.

\subsection{Protoplanet seed masses \& growth}
A model run begins with 100 seeds of $10^{-3} M_{\oplus}$ (roughly the mass of Pluto) placed logarithmically from 0.01 to 120 au. These seeds are placed simultaneously at t=0, although the model is capable of incorporating different formation times, mass, and composition of each seed.
We present an alternate model with a logarithmic distribution for the planetesimal seed masses and formation times in Appendix \ref{apx-planetesimal}.

The growth of a given seed is calculated by:

\begin{equation}
    \frac{dM}{dt}= f_{\text{peb}, \,t} \: \epsilon_t 
\end{equation}

where M is the seed mass, $f_{\text{peb}, t}$ and $\epsilon_t$ are the pebble flux and accretion efficiency at time \textit{t}, respectively. Growth is calculated per seed beginning from the outermost seed and progressing inward to account for filtering. Pebbles accreted by an outer seed are removed from the pebble flux and unavailable for inward seeds. The pebble flux is modified as:

\begin{equation}
    f_\text{{peb}, (i,\:t+1)}= f_\text{{peb}, (i,\:t+1)}\:  (1 - \epsilon_{o,\: t})
\end{equation}

where the subscripts i, o, are the inner and outer seed, respectively; the pebble flux for an inner seed is reduced by the fraction of pebbles accreted by an outer seed from the previous time step. Filtering is is cumulative, i.e., an inner protoplanet has the pebble flux reduced by the fraction accreted by each exterior protoplanet.

Accretion efficiencies vary with time and location (as more massive seeds accrete more efficiently). Efficiencies for a single seed remain <10\%  (usually <0.1\%) for much of the disk lifetime, and the filtering it performs is minimal. However, several dozen to a hundred seeds can affect the growth of inner seeds, particularly in delaying or preventing any particular protoplanet from reaching pebble isolation mass (see below).

Protoplanet seeds accrete pebbles until they reach pebble isolation mass at which point the protoplanet creates a pressure bump in the gas disk \cite{bitsch_pebble-isolation_2018}. This prevents pebbles from drifting inward, effectively quenching the supply and halting pebble accretion for all inner seed masses. Following \cite{lambrechts_forming_2014} and \cite{mulders_why_2021} we calculate the pebble isolation mass as:

\begin{equation}
    M_\text{{peb iso}} = 20_{M_{\oplus}} \:\bigg(\frac{M_\text{{star}}}{M_{\odot}}\bigg)  \bigg(\frac{H_\text{{gas}}}{0.05}\bigg)^3
\end{equation}

The pebble flux crossing a protoplanet that reaches the pebble isolation mass is reduced by a factor $10^3$, effectively quenching pebble drift.  \citet{bitsch_pebble-isolation_2018} theorize that sufficiently small grain sizes can nonetheless drift through the local gap created at the pebble isolation mass. However, alternate models where this factor is higher, or zero, did not affect the overall outcome.

\begin{table}
\caption{Disk properties and range.}
\label{table1}
\centering
\begin{tabular}{l c c c}
\hline\hline
Parameter & Unit & Value & Range  \\
\hline
$M_\text{{star}}$ & $M_{\odot}$ & 1.0 & 0.0625 - 2.0 \\
$f_\text{{disk}}$ &  & 0.08 & 0.01 - 0.453  \\
$f_{DG}$ &  &  0.0134  & \\
$\alpha$ &  & $10^{-3}$ &  \\
$r_{c}$ & au & 30  & \\
\hline 
\end{tabular}
\end{table}

\begin{table}
\caption{Planetesimal properties.}
\centering
\label{table2}
\begin{tabular}{l c c}
\hline\hline
Parameter & Unit & Value  \\
\hline
$n_\text{{seeds}}$ &  & 100  \\
$m_\text{{seeds}}$ & $M_{\oplus}$ & 0.001  \\
$r_\text{{seeds}}$ & au &  0.01 - 120\\
\hline 
\end{tabular}
\end{table}

\begin{table}
\caption{Snow line and changes from inside to outside the value defined by Equation \ref{RSL}.}
\centering
\label{table3}
\begin{tabular}{l c c c c}
\hline\hline
Parameter & Unit & &  & \\
\hline
$\rho$ & g/cm3 & 5.5 &$\rightarrow$ &2.5\\
ice:rock &  & 0 &$\rightarrow$&50/50  \\
$M_\text{{disk, solids}}$ &  &  0.5& $\rightarrow$ &1.0  \\
$v_\text{{frag}}$ & cm/s & 1 &$\rightarrow$ &10 \\
\hline 
\end{tabular}
\end{table}

\subsection{Modeling Strategy}
This introduction of The PPOLs Model presents a combination of pebble accretion mechanisms applied across a wide range of parameters. Tables \ref{table1} - \ref{table3} summarizes the important parameters, default values, and ranges. We present first a fiducial model using a 1$M_{\odot}$ star with $f_\text{{disk}} = 0.08$, a disk extending from 0.01 - 1000 au and alpha-viscosity $10^{-3}$.  The model runs for 10 Myr. One hundred planetesimal seed masses of $10^{-3}$ Earth mass are distributed logarithmically between 0.01 to 120 au. Next, we run a grid of models that vary in stellar mass and disk mass fraction to explore variations in planet formation outcomes and system architectures.

\section{Results}
First, we demonstrate that applying the mechanisms altogether (pebble filtering, pebble isolation mass, snow line evolution) yields an architecture distinct from considering each mechanism separately (Sect. \ref{three_mechs}). Next we vary only the disk mass fraction around a 1$M_{\odot}$ star (Sect. \ref{1msol}) and then by varying both the disk mass fraction and stellar mass (Sect. \ref{stellar_scale_mass}). Finally, we explore the consequences for water delivery to short period planets and habitable zone planets (Sect. \ref{watergrid}).

\begin{figure}
   \centering
   \includegraphics[width=0.7 \columnwidth]{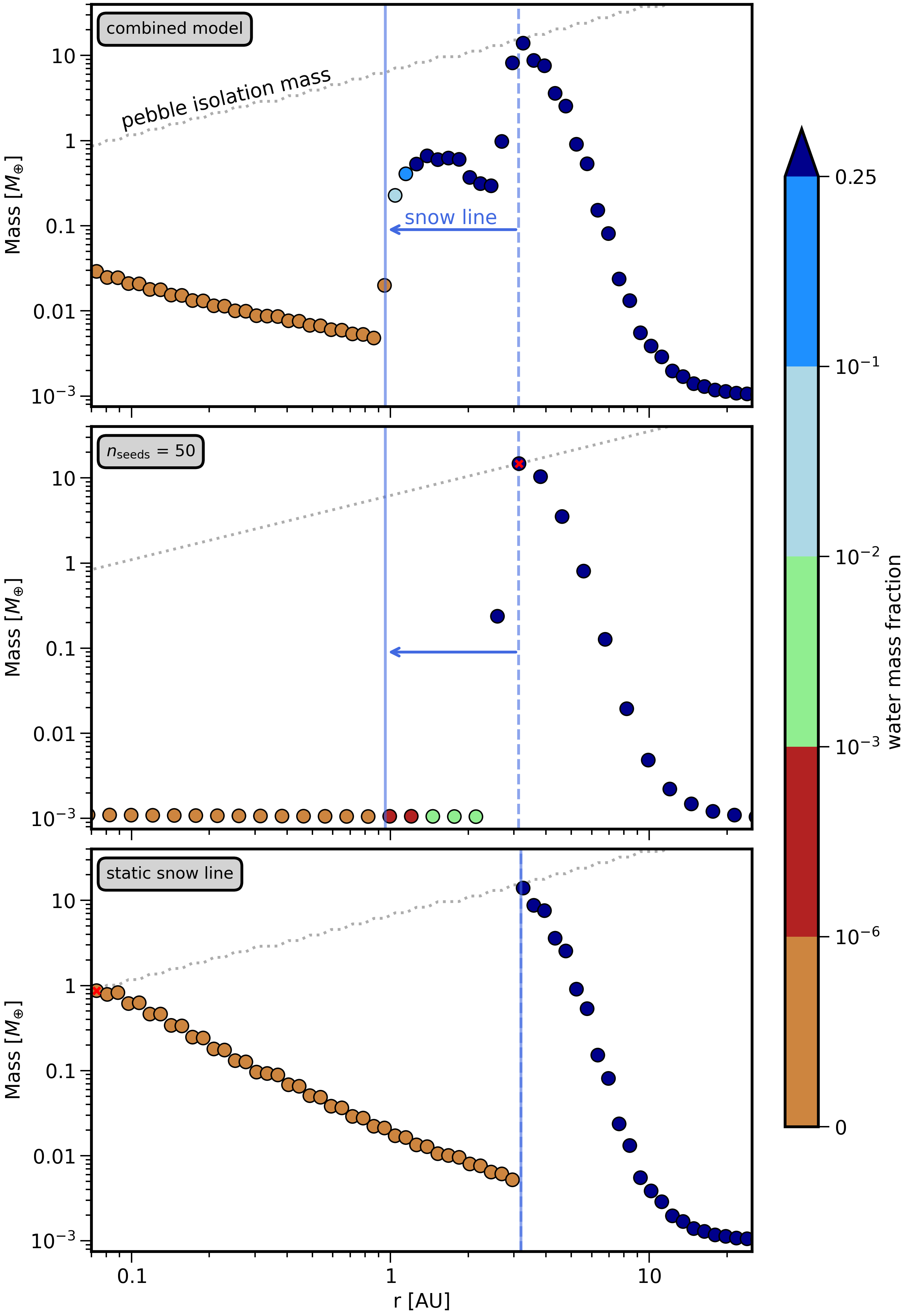}
      \caption{Protoplanet systems from three models of the combined mechanisms (top), fewer seeds (middle), and a static snow line (bottom). Seed masses that reach the pebble isolation mass (grey dashed line) are marked (red X). The architecture of the combined mechanisms yields a clear difference from isolated mechanisms.}
      \label{FigThreeMechs}
   \end{figure}

   \begin{figure}
    \centering
    \includegraphics[width=0.7\columnwidth]{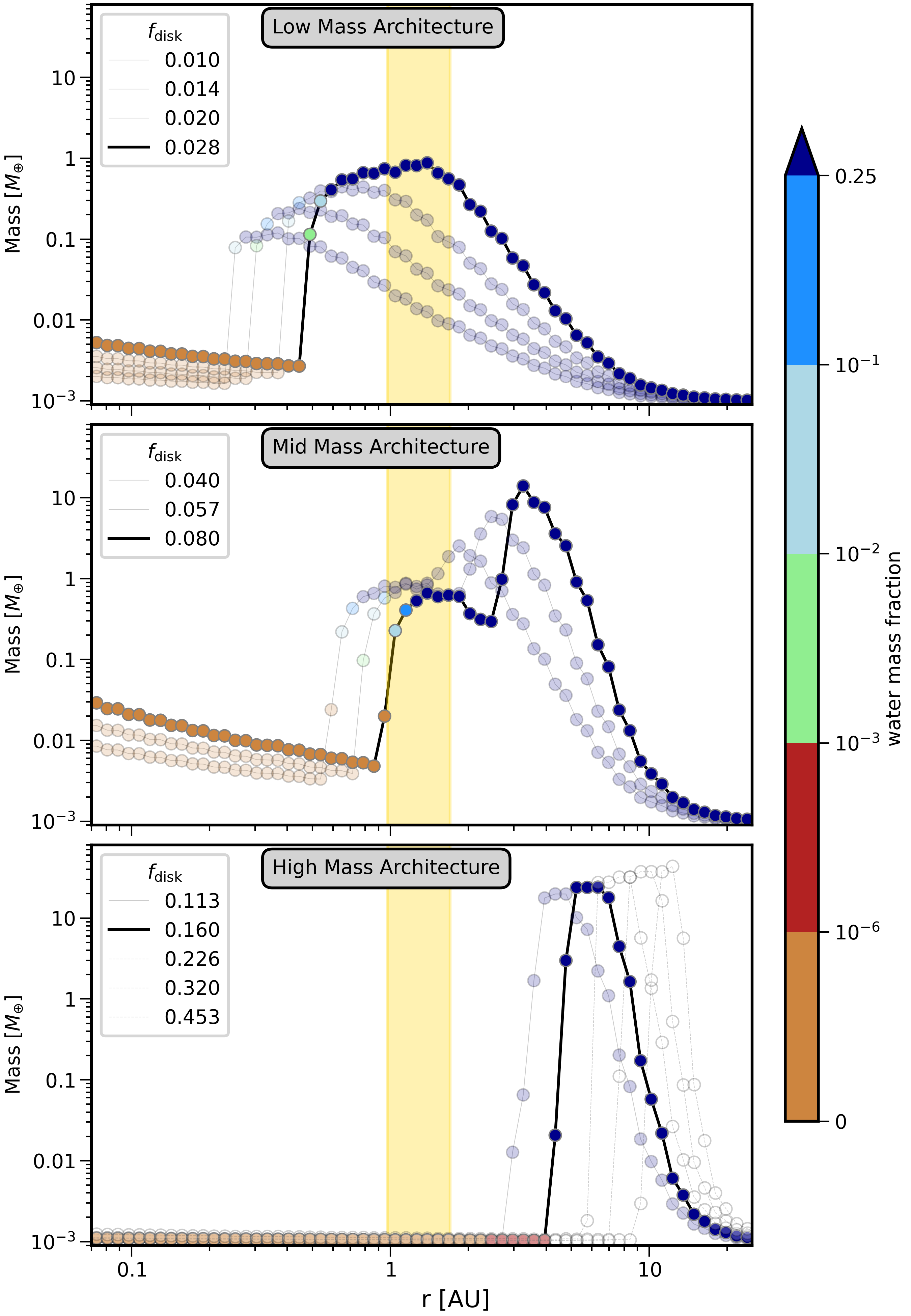}
    \caption{Resulting planetary systems around a 1$M_{\odot}$ star separated into three architectures based on disk mass. Gold-shaded region corresponds to the conservative habitable zone from \cite{kopparapu_habitable_2014}. Higher disk mass fractions that are unstable are dashed lines.}
    \label{FigDMFs}
\end{figure}

\subsection{Combined effects of filtering and snow line evolution in single model} \label{three_mechs}
The combined effects of filtering from many seeds extending into the outer disk and the snow line evolution yields a distinct system architecture compared to models with only one of these mechanisms. 

The combined mechanisms (Fig. \ref{FigThreeMechs}, top panel) enables protoplanets swept over by the snow line to grow orders of magnitude larger and hydrated compared to one mechanism at a time (Fig. \ref{FigThreeMechs}, middle and bottom panels). The snow line separates rocky from icy pebbles, and we assume icy pebbles have a higher fragmentation velocity and survive to larger typical grain sizes than their rocky counterparts in the inner disk. This increases the relative pebble accretion efficiency for icy pebbles, and therefore boosts growth to the seeds the snow line sweeps over. Pebble filtering from more seeds in the outer disk prevents the first seed past the snow line from reaching pebble isolation mass and allows growth to proceed at all. The combined model yields protoplanet masses distributed in a double-peaked "humpback" with water mass fractions that decreases towards the inner disk and trace the evolution of the snow line. 

The model with fewer seeds (Fig. \ref{FigThreeMechs}, middle panel) experiences less efficient pebble filtering, and the first seed past the snow line reaches pebble isolation mass; this quenches the icy pebbles drifting to the inner disk even as the snow line continues to evolve. The pebble flux remains limited inside the snow line and therefore growth in the inner disk is also limited.

The model with the static snow line (Fig. \ref{FigThreeMechs}, bottom panel) experiences as much pebble filtering as the combined model, and so the first seed past the snow line is prevented from reaching pebble isolation mass. Therefore, the pebble flux to the inner disk remains high and dry as icy pebbles sublimate. Interior to the snow line, protoplanet growth proceeds from the inside out, typical of pebble accretion models \citep[e.g.][]{lambrechts_forming_2014}. A net outcome of the static snow line is to shift growth to the inner disk so that several inner seeds near 0.1 au reach their pebble isolation mass (0.1 - 1$M_{\oplus}$) while seeds near 1 au  remain less massive (0.01 - 0.1$M_{\oplus}$). The peak's location near 0.1 au is a consequence of the gas disk lifetime: truncating the model at 6 Myr instead of 10 Myr shifts the peak inward and downward (not shown).

\subsection{Architectures based on disk mass} \label{1msol}
Next we maintain the stellar mass at 1$M_{\odot}$ and run the combined pebble accretion mechanisms for a range of disk mass fractions from 0.01 to 0.45, corresponding to 45 - 2000$M_{\oplus}$ of initial solids. We apply this wide range of disk mass fractions to encompass all outcomes and reveal edge-case behavior. The Toomre criterion reaches Q < 2 for all stellar masses with $f_\text{{disk}}$ = 0.453. Models with 1 and 2 $M_{\odot}$ stars are unstable (Q < 2) for $f_\text{{disk}}$ > 0.226. We separate the outcomes into a low-, medium-, and high-disk mass architecture (Fig. \ref{FigDMFs}) based on outcomes that yield the double-peaked "humpback" mass distribution (medium-mass), and those disk mass fraction models below and above that range. Specifically, the high-mass architecture occurs when the first seed past the snow line is able to reach pebble isolation mass.

The low-mass architecture (Fig.\ref{FigDMFs}, top panel, $f_\text{{disk}}$ $>$0.03) yields mostly water-rich embryos at close separations from the host star, and a smaller number of embryos between Mars and Earth mass with mixed water mass fractions down to $10^{-2}$ - $10^{-3}$ . The snow line position is set by the initial gas surface density ($\Sigma_{0,g}$), which in turn is set by the total disk mass. Therefore, with lower disk mass, the initial snow line position is closer, so icy pebbles are able to accrete directly onto seeds down to ${\sim}$0.3 au. The dust surface density is too low overall to sustain a high pebble flux for both dry and icy pebbles, and therefore the highest mass protoplanet is 1$M_{\oplus}$ at 1-2 au, with virtually no growth beyond 10 au. The habitable zone is beyond the snow line range, and all protoplanets in the habitable zone become water rich (water mass fraction = 0.5).

The medium mass architecture (Fig. \ref{FigDMFs}, middle panel, 0.04 < $f_\text{{disk}}$ < 0.08) yields both large outer protoplanet cores (>2.5 $M_{\oplus}$) and Earth-mass inner cores in the same system, distinguishing it from the other architectures. Seeds swept by the snow line benefit from higher growth efficiency via icy pebbles, and they in turn filter the mass from reaching the inner seeds. The dust surface density is higher and supports a higher overall pebble flux, particularly past 1 - 2 au. The snow line is pushed further out and enables 1) efficient growth of protoplanets beyond 3 au and 2) fine amounts of water delivered to Earth-mass embryos in the habitable zone. Unlike the lower and higher mass architectures, here the snow line evolution sweeps through the habitable zone for 1$M_{\odot}$.

The high mass architecture (Fig. \ref{FigDMFs}, bottom, $f_\text{{disk}}$ > 0.11) is defined by outer seeds past the snow line efficiently reaching pebble isolation mass regardless of pebble filtering. This architecture produces water-rich cores between 1 - 10+ $M_{\oplus}$ at 5 au  with water-rich cores beyond them spanning orders of magnitude smaller. Beyond 15-20 au, the disk surface density decreases and pebble accretion is inefficient for seeds of our intial Plutonian mass. Classical estimation for the onset of rapid gas accretion occurs at core masses ${\sim}$10 $M_{\oplus}$ \citep{mizuno_formation_1980, pollack_formation_1996}. These higher mass models clear that threshold, and it is likely these cores would accrete significant gas envelopes.

\subsection{Architectures scale with stellar mass} \label{stellar_scale_mass}
\begin{figure}
    \centering
    \includegraphics[width=0.7\columnwidth]{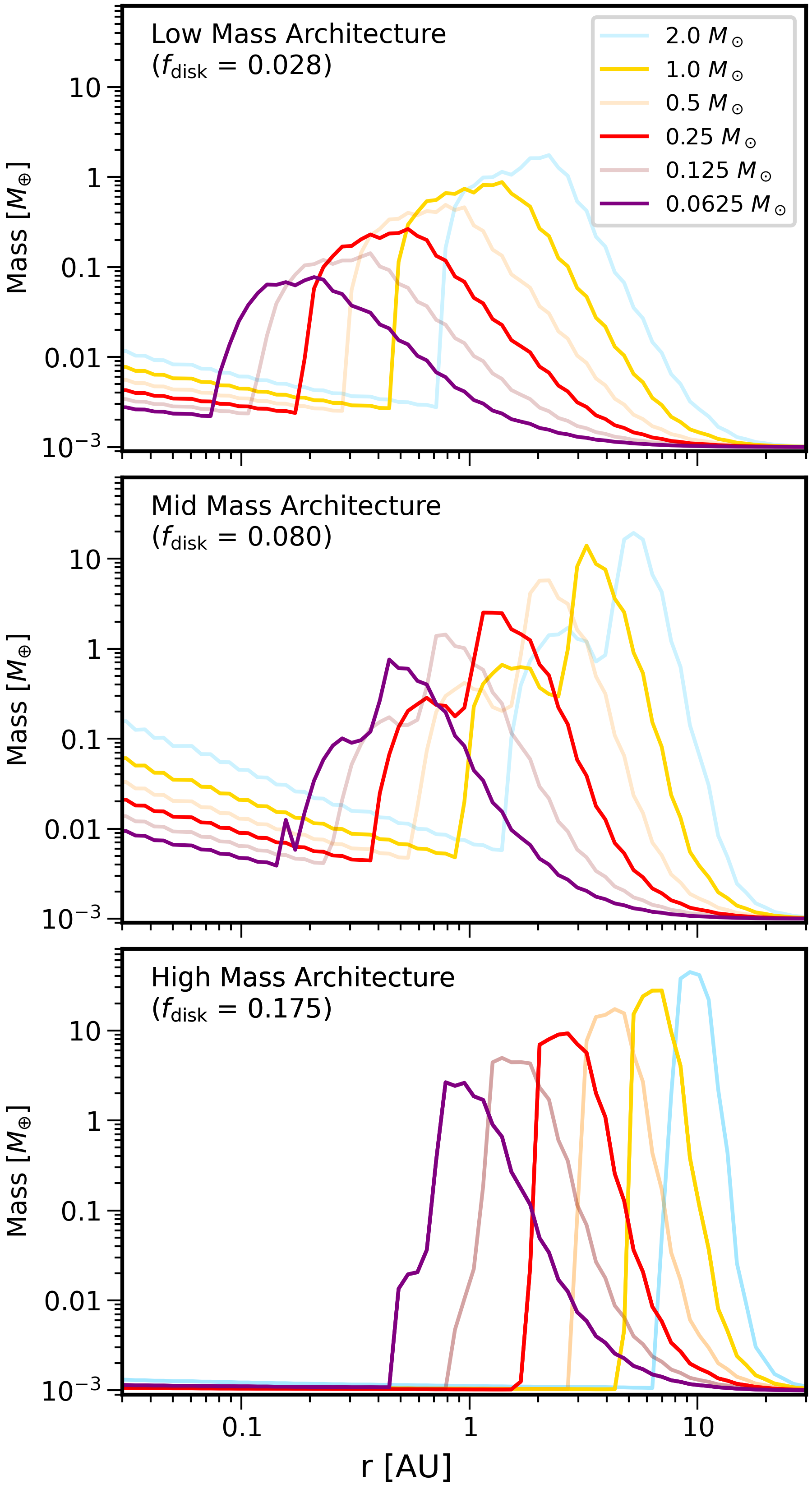}
        \caption{Planetary architectures for all stellar masses used at three specific disk mass fractions. For clarity, individual protoplanet masses are not shown and alternate stellar mass models are faded. The split in architecture is consistent for a particular disk mass across different stellar mass. The representative high-mass architecture satisfies Q > 2 for all models shown.}
        \label{FigAllMSOL}
   \end{figure}

We ran our pebble accretion model with a representative low, medium, and high-disk mass fraction (0.028, 0.08, 0.226) for a range of stellar mass from 0.0625 - 2$M_{\odot}$, encompassing low mass M dwarfs to A-type stars. Figure \ref{FigAllMSOL} shows the different architectures based on disk mass fraction remain even when stellar mass varies, emphasizing the significance of disk mass in influencing outcome architectures.

In the low-mass architecture no embryo emerges above ${\sim}$3$M_{\oplus}$, and the majority are < 1$M_{\oplus}$. It is unlikely that pebble accretion can produce a proto-Jupiter core in the low-mass architecture, regardless of stellar mass. However, these models use a single metallicity throughout the disk - and it is possible that higher metallicity may enable giant planet cores to form by simple virtue of more solids in the disk \citep{swain_planet_2024}. The protoplanet mass is distributed as a single plateau with relatively uniform mass for a given stellar mass; for the lowest mass stars, protoplanet masses reach Mars-mass between 0.1-0.3 au, while the highest stellar mass model contains protoplanets around an Earth mass from 1-3 au.

In the high-mass architecture, ${\sim}$10$M_{\oplus}$ embryos (gas giant cores) appear for stars $\geq$ 0.25$M_{\odot}$, and less massive stars are still able to produce large ${\sim}$3-5$M_{\oplus}$ cores, consistent with becoming super-Earths or mini-Neptunes.

The mid-mass mass architecture is more complex: each stellar mass yields the bimodal mass distribution definitive of this regime,  but overall protoplanet mass scales with stellar mass.  Generally the protoplanet mass of the outer peak is an order of magnitude larger than then inner peak. For models with 1$M_{\odot}$ this translates to an inner peak of 1$M_{\oplus}$ near 1 au and outer peak of 10$M_{\oplus}$ near 4-5 au. For stars <0.5$M_{\odot}$, the inner peak is roughly Mars mass and outer core is no more than 1 - 2$M_{\oplus}$. The outer peak contains water-rich protoplanets because they formed and accreted material entirely beyond the snow line, while the inner peak protoplanets can contain varying levels of water, down to <1\% water. 

\subsection{$H_2O$ across the grid of stellar mass and disk mass} \label{watergrid}
\begin{figure}
    \centering
    \includegraphics[width=0.7\columnwidth]{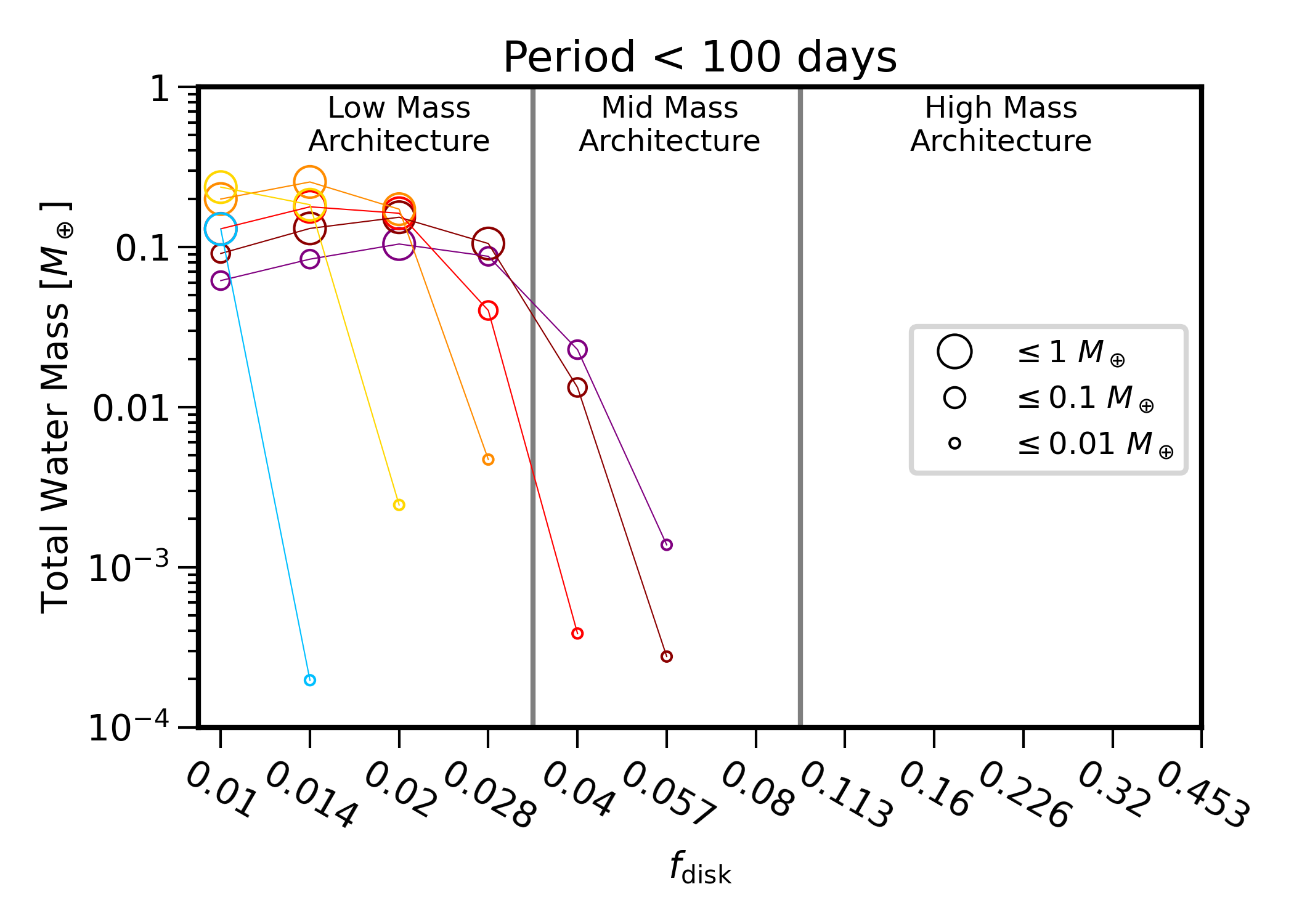}
    \caption{Sum of water mass accreted by planets with P < 100 days, by stellar mass (color lines) and disk mass fraction along the x-axis. Circle size represents the average protoplanet mass with P<100 days in each model. Low-mass architectures across all stellar mass receive the highest amounts of water distributed among protoplanets between Mars-Earth mass, while the high-mass architecture is unable to deliver water to such inner protoplanets.}
    \label{fig:FigPlot100dayWater}
\end{figure}

\begin{figure}
    \centering
    \includegraphics[width=0.7\linewidth]{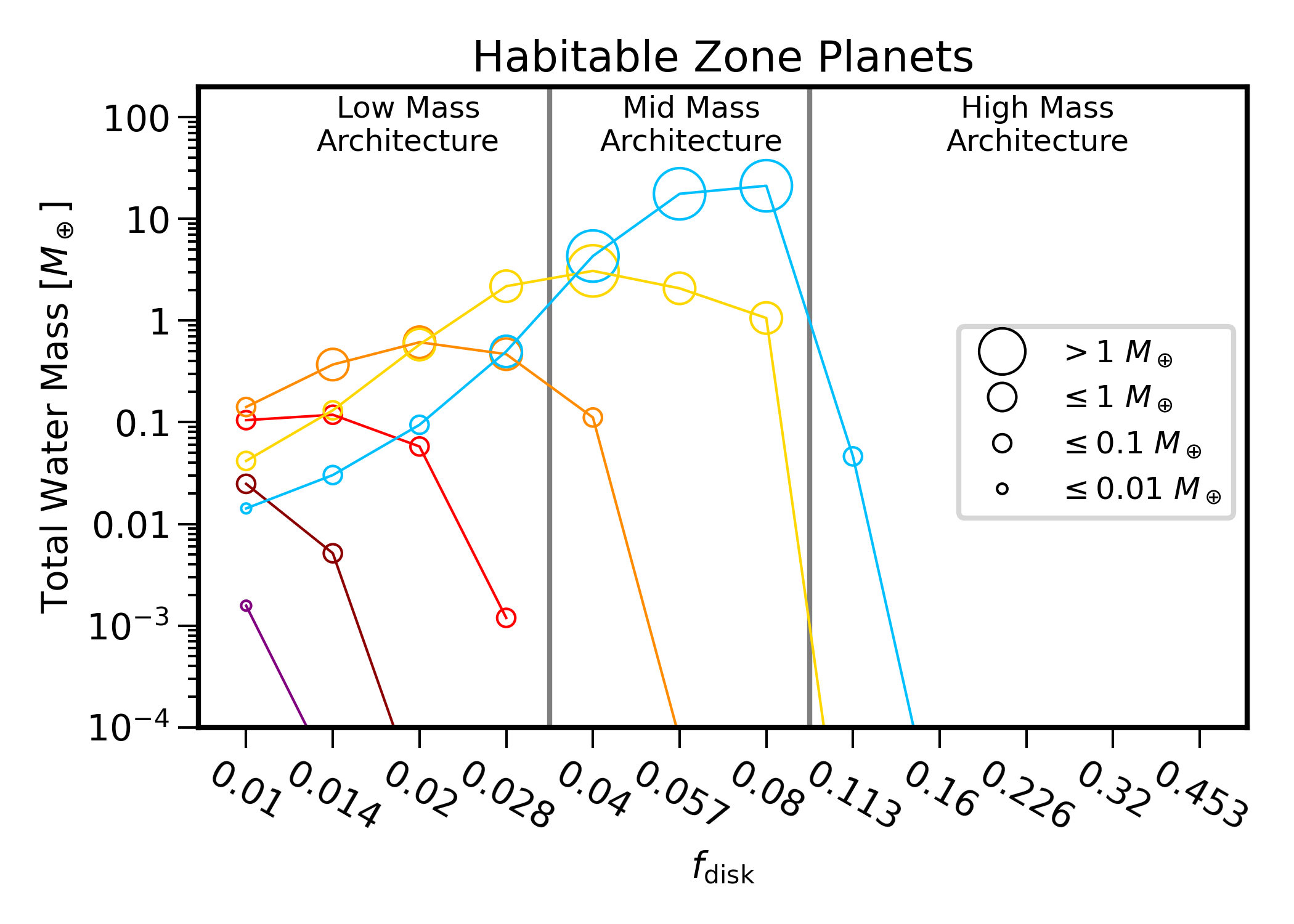}
    \caption{Sum of water mass accreted by planets in the habitable zone of their respective stars (color lines) and disk mass fraction along the x-axis. Circle size represents average protoplanet mass among each habitable zone containing 6 to 8 protoplanets. The highest mass protoplanets are also the most water-rich and occur in the middle-mass architecture, while the high-mass architecture is unable to deliver water to protoplanets in the habitable zone (truncated from the plot are models where total water mass < $10^{-8} M_{\oplus}$). The low-mass architecture features a mix of water and protoplanet mass <1$M_{\oplus}$.}
    \label{fig:FigPlotHZwater}
\end{figure}

We summarize the delivery of icy pebbles into protoplanets for all stellar/disk mass combinations of this work in the context of disk mass-based architectures. Specifically, we present the total water mass contained in protoplanets in two regions of interest: for protoplanets with P < 100 days (Fig. \ref{fig:FigPlot100dayWater}) and within the conservative habitable zone (HZ) of their star (Fig. \ref{fig:FigPlotHZwater}).

Water can accrete into short period protoplanets for all stellar mass only in the low-mass architecture and for stars >0.25$M_{\odot}$ in the mid-mass architecture. At the lower disk mass, the snow line exists closer to the star and can reach these short separations.  In contrast, the snow line range for middle- and high-mass architectures is beyond 100-day periods. Further, the low disk mass starves the outer protoplanets overall from reaching pebble isolation mass and keeps the pebbles flowing. 

Delivering icy pebbles to a star's habitable zone requires the snow line evolution (determined by disk conditions during formation) to overlap with the habitable zone (determined by insolation and a planet's atmosphere). As the disk mass fraction increases, the snow line range is pushed to greater distances from the star and allows for more overlap with the habitable zone \citep[see Fig. 1 in][]{mulders_snow_2015}. Therefore, water mass in a star's habitable zone will peak as there is more overlap between the snow line evolution and the HZ. Despite the small difference in the number of habitable zone protoplanets for more massive stars (6) compared to low mass stars (8), generally more water is accreted in the HZ for more massive stars.

For increasing disk mass, the first seed beyond the initial snow line is able to grow to pebble isolation mass and quench growth inward. In theory, some icy pebbles may still coagulate  inside this distance as the snow line evolves inward. However, the (now quenched) pebble flux is overall too low to drive significant or efficient growth. This explains both the vanishing amount of water accreted to protoplanets in the HZ that are not shown in Fig. \ref{fig:FigPlotHZwater}, as well as the sharp decline for 1 and 2$M_{\odot}$ stars.

\section{Discussion} \label{discussion}
It remains a mystery (for now) how common true solar-system analogs are among the exoplanet population, currently limited by observational capabilities. It may be that the fraction of systems that form both archetypal Earths and gas giants in the same system is small because the star/disk combinations that can produces them is a small fraction of greater natural variability in disk masses. We qualitatively suggest that the middle-mass architecture resembles an early Solar System architecture: outer cores grown efficiently enough during the gas disk lifetime with some growth and water delivery to inner rocky cores. The prevalence of compact terrestrial mass planets - so-called "peas in a pod" architecture typical of multi-planet systems discovered by Kepler may arise primarily from lower disk masses. The architectures presented here arise from the gas-disk phase of planet formation that does not include migration; further, it is expected that a prolonged collisional phase follows the stage of planet formation studied here. Therefore, although the middle-mass architectures bear protoplanets that are water-rich and massive, there remains theoretical room for dynamical scenarios like the Grand Tack \citep{walsh_low_2011, raymond_grand_2014} to further sculpt the system. 

The total water mass in regions of interest (Figs. \ref{fig:FigPlot100dayWater} \& \ref{fig:FigPlotHZwater}) is useful to set expectations for the bulk transport and delivery icy pebble mass in a protoplanetary disk, and these mechanisms may also be relevant for water vapor. \cite{easterwood_water_2024} analyze JWST spectra of small protoplanetary disks and find more water vapor in the inner disk region than their wider counterparts. They speculate that larger disks may be more likely to exhibit a mechanism that halts the inward flux of icy material - a notion attributed to some by planet formation, rings, or other gap-forming scenarios \citep{kalyaan_effect_2023}. Here, identifying the broad expectations for icy pebble deliver into the inner disk halted by more massive disks is consistent with this emerging picture.

The PPOLs Model uses several simplifying assumptions, among the boldest is a lack of gas accretion and evolution on to the central star. Several effects are likely to arise by considering such gas evolution. One is the gas temperature profile takes on a viscous-to-irradiation regime \citep[e.g.][]{ida_radial_2016}, which affects the broad properties of the disk like the scale height, which in turn affect the pebble isolation mass at a given part of the disk. Further, the snow line may be placed and evolve differently. In our parametrization (Eq. \ref{RSL}), the snow line is explicitly set by the dust/gas fraction and evolves inward as the dust decreases. If the gas also decreases, the snow line may not evolve as far inward. However, decreasing gas density may also have confounding secondary effects on the ability for dust to coagulate into dust, or drift. Nonetheless, our parametrization is self-consistent within the context of the model.

One assumption of this work is the difference in fragmentation velocity between rocky and icy pebbles, which is supported by laboratory experiments by \cite{gundlach_stickiness_2015} and \cite{schrapler_collisional_2022}. However, other recent experiments by \cite{musiolik_contacts_2019} describe that water ice does not significantly stick or grow over rocky particles. \cite{drazkowska_planet_2023} emphasize further than the fragmentation threshold may be more dependent on local disk conditions and ultimately remain largely unexplored. Exploring outcomes of different or equal fragmentation velocity \citep[as done by e.g.][]{chachan_small_2023} between rocky and icy pebbles is a topic for future work.

We do not consider gas- or dust-driven migration in our model, in line with the simplified treatment of the gas dynamics. Gas-driven migration suggests these protoplanets would migrate inward once they reach ~Mars mass and higher, while the effects of dust-driven migration are less clear. Our results present an overall lack of protoplanets $>10^{-3} M_{\oplus}$ within 0.1 au and suggests inward migration is necessary to explain such observed systems. Alternatively, the scattering and redistribution of mass in the late-stage collisional growth phase could provide the conditions to assemble such rocky planets and hydrate them.

\section{Conclusions}
We present the PPOLs Model for producing planetary systems grown via the pebble snow mechanism. By developing upon previous models of pebble accretion, we combine important mechanisms usually analyzed individually that sculpt the overall mass and water content of planetary systems. We summarize key insights:

\begin{enumerate}
    \item The cumulative filtering of seed masses distributed into the outer disk is sufficient to prevent key inner seeds from reaching the pebble isolation mass.
    \item A snow line that evolves inward will shift protoplanet growth from 0.1 au outward to the snow line and enable delivery of icy pebbles to inner seeds.
    \item Cumulative filtering and an evolving snow line together yield a distinct system architecture compared to considering each single mechanism separately.
    \item When applied to different disk mass fractions and stellar mass, the disk mass fraction produces three different architectures via pebble snow:
    \begin{itemize}
        \item A low-mass disk yields numerous cores between Mars and Earth mass, with a mix of water mass fractions, interior to the star's habitable zone.
        \item A high-mass disk yields numerous water-rich cores capable of accreting significant gas envelopes beyond the habitable zone of their stars. These cores reach isolation mass and prevent pebble accretion for inner cores.
        \item A middle-mass disk yields both inner Earth-mass cores and outer pre-giant planet cores, with Earth-mass cores containing decreasing water mass fractions, approaching Earth-like amounts in the habitable zone of solar mass stars.
    \end{itemize}
\end{enumerate}

The PPOLs Model does not produce the final planets of the Solar System, or indeed any system, outright. The outputs represent distributions of protoplanets at the transition from the gas-disk to the prolonged, late-stage collisional growth phase, and so remain incomplete. Finishing the planet formation process requires n-body simulations to explore the final assemblage of the embryos into the "last planets standing". The PPOLs Model provides reasonable inputs to such simulations of embryo masses and water mass fractions that have been derived from pebble accretion in a self-consistent way. Using the outputs of these models as inputs to n-body simulations is a natural next step future work. Connecting mechanisms of planet formation to the resulting architecture will help parse through the observed diversity of exoplanets and assist in targeted searches for true Solar System analogs and broader patterns in planetary systems.

\begin{acknowledgements}
S.M. expresses the highest gratitude for support from the North Dakota Space Grant Consortium. G.D.M. acknowledges support from FONDECYT project 1252141 and the ANID BASAL project FB210003. This material is based upon work supported by the National Aeronautics and Space Administration (NASA) under Agreement No. 80NSSC21K0593 for the program “Alien Earths”. The results reported herein benefited from collaborations and/or information exchange within NASA’s Nexus for Exoplanet System Science (NExSS) research coordination network sponsored by NASA’s Science Mission Directorate. This material is based upon work supported by the National Aeronautics and Space Administration under Grant No. 80NSSC20M0102 issued through the Space Grant College and Fellowship Project, North Dakota Space Grant Consortium. Any opinions, findings, conclusions, or recommendations expressed in this material are those of the author and do not necessarily reflect the views of NASA.
\end{acknowledgements}

\newpage
\appendix

\section{Temperature Profile}\label{apx-temp}
The biggest difference between the power law used in the main text and the profile suggested in \citet{ida_radial_2016} is the irradiation regime is cooler into the outer disk. This lowers the pebble isolation mass by affecting the gas scale height in our model and in turn renders it easier for a seed mass to reach that threshold and quench the pebble flux. We explore the impact of this difference by running models for a range of stellar mass with the viscous/irradiation heating profile of \citet{ida_radial_2016}, with disk masses that span to lower values (down to $10^{-3}$ instead of $10^{-2}$). Specifically, we use
\begin{equation}
    T_\text{vis} = 200 \: \frac{M_\text{star}}{M_{\odot}}^{3/10} \: \frac{\alpha}{10^{-3}}^{-1/5} \frac{\dot M}{10^{-8}M_{\odot}/\text{yr}} \: \bigg(\frac{r}{1\:\text{au}}\bigg)^{-9/10},
\end{equation}
and
\begin{equation}
    T_\text{irr} = 150 \: L_{\odot}^{2/7} \: \frac{M_\text{star}}{M_{\odot}}^{-1/7} \: \bigg(\frac{r}{1\:\text{au}}\bigg)^{-3/7} ,
\end{equation}

\begin{figure}
\centering
\includegraphics[width=0.65\columnwidth]{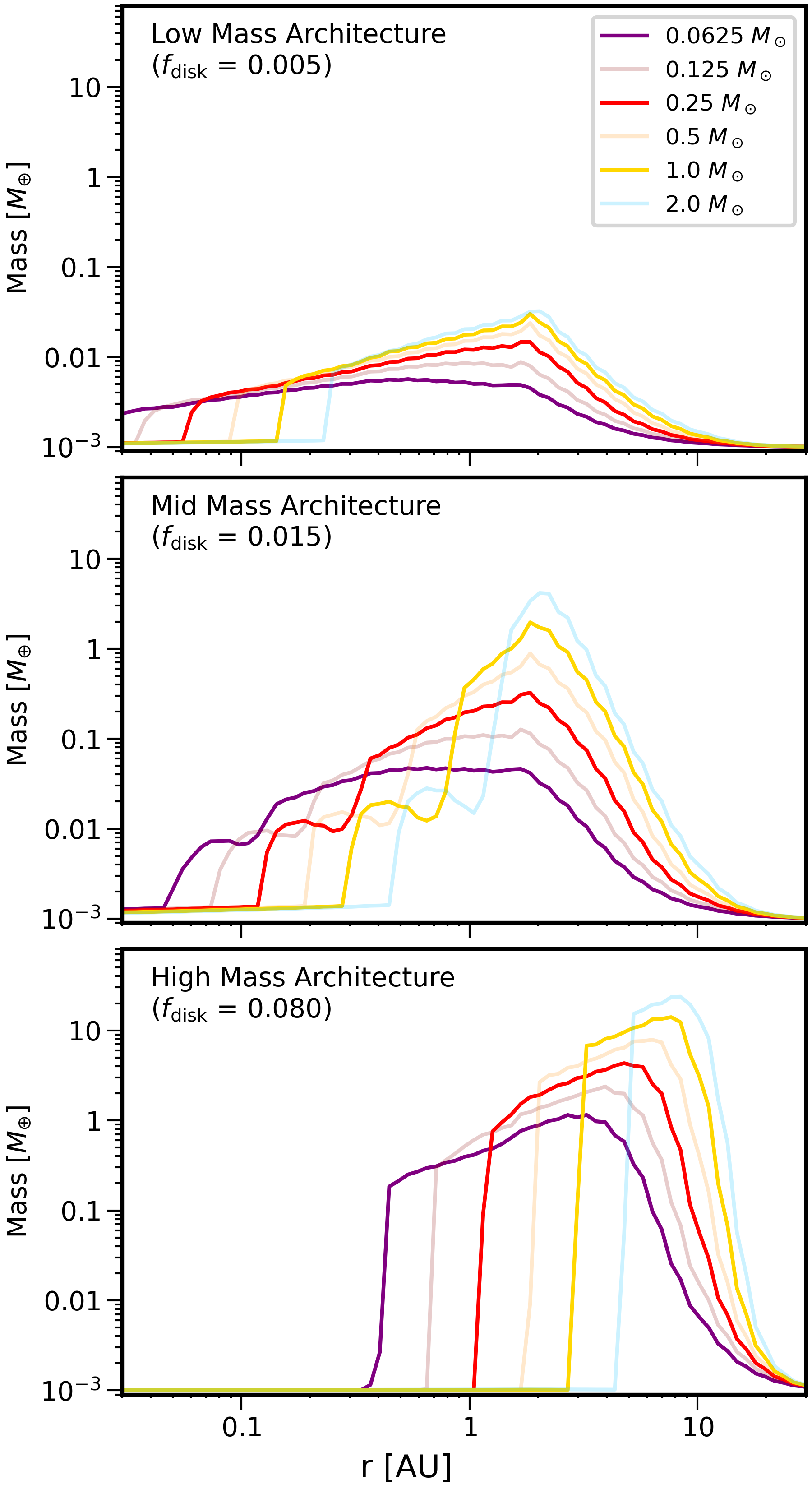}
    \caption{Analogous to Figure \ref{FigAllMSOL} showing lower disk mass fractions and the viscous/irradiation temperature profile for all stellar masses. For clarity, individual protoplanet masses are not shown and alternate stellar mass models are faded. The split in architecture is consistent for a particular disk mass across different stellar mass.}
    \label{FigApx-alttemp}
\end{figure}

where $\dot M$ is the gas accretion rate and $L_{\odot}$ is the stellar luminosity relative to the solar value. We assume the gas accretion rate scales as $M^2$\citep{manara_demographics_2023}. We find that similar architectural blueprints emerge (Fig. \ref{FigApx-alttemp}, albeit scaled down to lower overall protoplanet masses, particularly for the inner mass peak of the "middle-mass" architecture. Considering the limiting factor is the smaller pebble isolation mass, it is possible that disk temperatures increase from the irradiation regime closer to the observed power law as the dust depletes. If so, the pebble isolation mass likely also increases, and realized outcomes may resemble those for the power law temperature profile.

\section{alpha-viscosity}\label{apx-alpha}
\begin{figure}
\centering
\includegraphics[width=0.95 \linewidth]{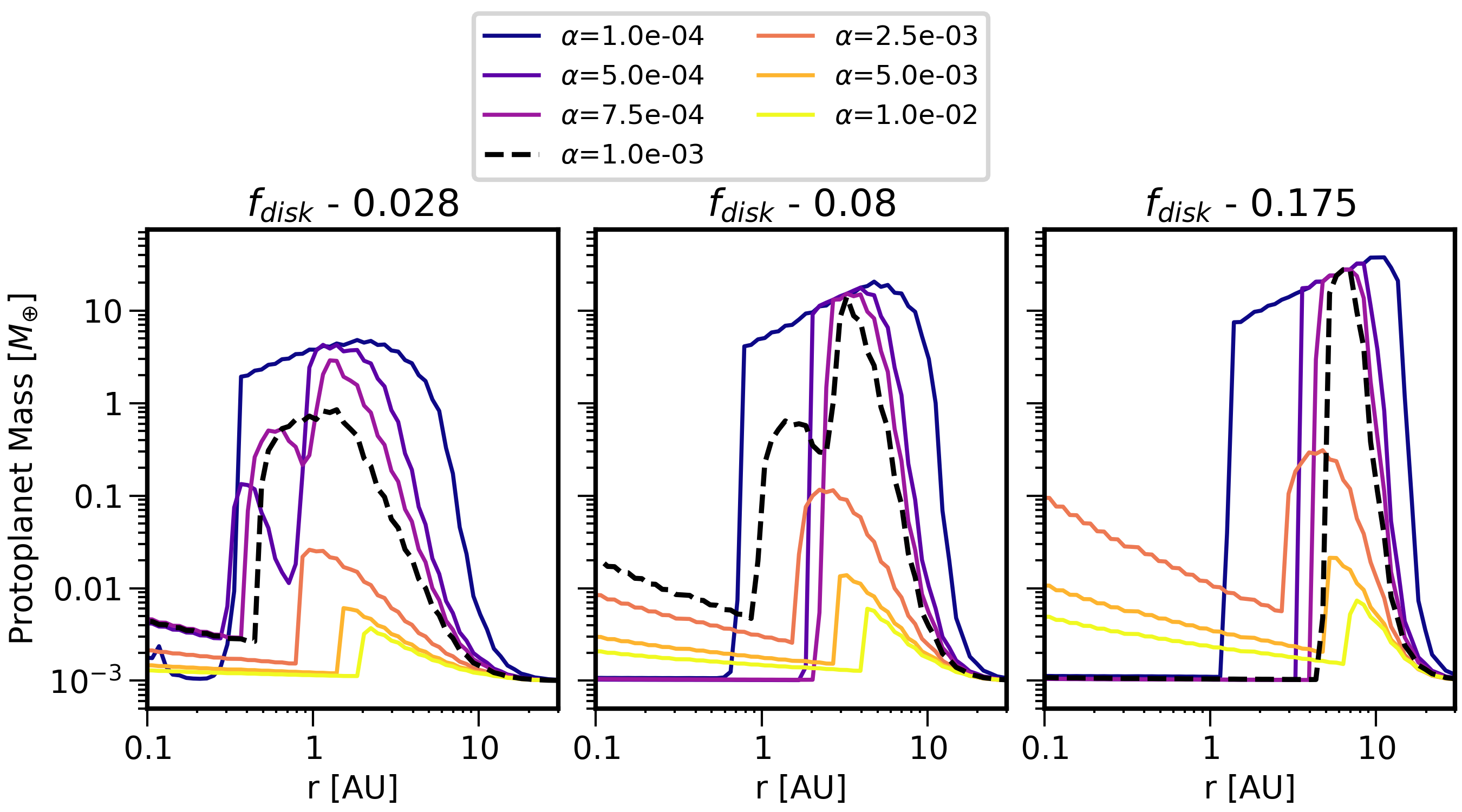}
    \caption{Protoplanet masses around 1 $M_{\odot}$ star for a typical low-, medium-, and high-disk mass for a range of alpha-viscosity values. The default value ($10^{-3}$) is black, while lower values are purple and higher values are yellow.}
    \label{FigApx-alphas}
\end{figure}

We perform models for 1 $M_{\odot}$ star at the same low-, medium, and high-disk masses from Figure \ref{FigDMFs} using alpha-viscosity values from $10^{-4} - 10^{-2}$ (Fig. \ref{FigApx-alphas}). Lower alpha makes accretion more efficient, and reproduces the “high-mass” architecture; higher alpha stifles accretion and drives the outcome towards the “low-mass” architecture. With lower values of alpha, it is possible to reproduce the interesting twin-peak architecture at a lower disk masses.

\section{Planetesimal Distribution}\label{apx-planetesimal}
Classical estimation suggests that planetesimals, and thus the seed masses for this model, form from the "inside-out" of the disk and may increase in mass at greater orbital distances. We explore the effect on the outcome by growing these seeds with increasing mass ($10^{-5}$ – $10^{-2}  M_{\oplus}$) and formation time ($10^{-2}$ – $5 \times10^{5}$ yrs) with orbital distance. With this distribution, seeds at 1 au are introduced with mass $4\times10^{-4}$ $M_{\oplus}$ at $10^{4}$ yrs, and at 5 au with $9\times10^{-4}$ $M_{\oplus}$ at $4\times10^{4}$ yrs. This is consistent with growth curves presented in \citet{kobayashi_planetary_2010} (see Figures 7 and 9) via the initial runaway phase of collisional embryo growth.

\begin{figure}
\centering
\includegraphics[width=0.75 \linewidth]{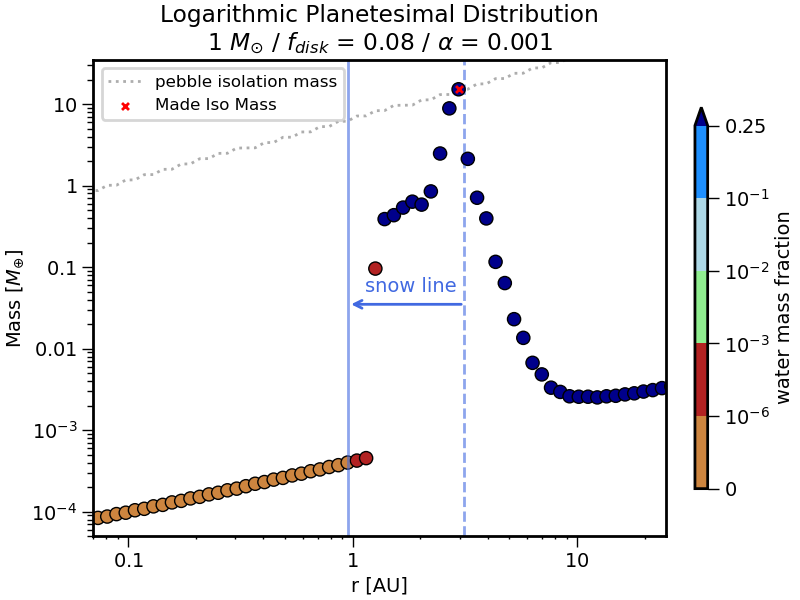}
    \caption{Protoplanet masses around 1 $M_{\odot}$ star using a logarithmic distribution of planetesimal seed mass and introduction time, with the same disk parameters as Fig. \ref{FigDMFs}, middle panel. The later introduction of outer seeds do not filter the inward pebble flux enough to prevent the seed mass at 3 au from reaching pebble isolation mass (red 'x').}
    \label{FigApx-planetesimallog}
\end{figure}

For a $1 M_{\odot} star$, $f_\text{disk}$=0.08, and $\alpha=10^{-3}$, the model outcomes are largely consistent with the uniform distribution, provided the seed masses appear within ${\sim}6$ au within ${\sim}10^{5}$ years. The later introduction of seeds with distance do not filter the pebble flux as much as the uniform distribution, and seeds can reach the pebble isolation mass (Fig. \ref{FigApx-planetesimallog}).

\clearpage
\bibliography{mypapercitations}{}
\bibliographystyle{aasjournalv7}
\end{document}